%% file: main.tex
\documentclass[letterpaper, 10 pt, conference]{ieeeconf}  

\IEEEoverridecommandlockouts                              
\title{\LARGE \bf
Privacy-Engineered Value Decomposition Networks for Cooperative Multi-Agent Reinforcement Learning
}

\author{Parham Gohari, Matthew Hale, and Ufuk Topcu
\thanks{P. Gohari is with the Department of Electrical and Computer Engineering, and Robert Strauss Center for International Security and Law, University of Texas at Austin,
        Austin, TX, USA
        {\tt\small pgohari@utexas.edu}}
\thanks{M. Hale is with the Faculty of Mechanical and Aerospace Engineering, University of Florida, Gainesville, FL, USA
        {\tt\small matthewhale@ufl.edu}}
\thanks{U. Topcu is with the Faculty of Aerospace Engineering and Engineering Mechanics, University of Texas at Austin, Austin, TX, USA
        {\tt\small utopcu@utexas.edu}}%
}

\newcommand{\policy}{\pi}
\newcommand{\jpolicy}{\boldsymbol{\pi}}

\newcommand{\jaction}{\boldsymbol{a}}

\newcommand{\jtraj}{\boldsymbol{\tau}}

\newcommand{\localQ}{\{Q_i\}_{i \in \mathcal{N}}}
\newcommand{\bs}[1]{\boldsymbol{#1}}
\usepackage{amsmath, amssymb}
\usepackage{pgfplots}

\usepackage{algorithmic}
\usepackage{amsfonts}
\usepackage[ruled, noend, linesnumbered, vlined]{algorithm2e}
\usepackage{hyperref}
\hypersetup{
    colorlinks=true,
    linkcolor=blue,
    filecolor=magenta,      
    urlcolor=cyan,
    pdftitle={Overleaf Example},
    pdfpagemode=FullScreen,
    }

\newtheorem{definition}{Definition}
\newtheorem{theorem}{Theorem}

\begin{document}

\maketitle
\thispagestyle{empty}
\pagestyle{empty}

\begin{abstract}
In cooperative multi-agent reinforcement learning (Co-MARL), a team of agents must jointly optimize the team's long-term rewards to learn a designated task. Optimizing rewards as a team often requires inter-agent communication and data sharing, leading to potential privacy implications. We assume privacy considerations prohibit the agents from sharing their environment interaction data. Accordingly, we propose Privacy-Engineered Value Decomposition Networks (PE-VDN), a Co-MARL algorithm that models multi-agent coordination while provably safeguarding the confidentiality of the agents' environment interaction data. We integrate three privacy-engineering techniques to redesign the data flows of the VDN algorithm—an existing Co-MARL algorithm that consolidates the agents' environment interaction data to train a central controller that models multi-agent coordination—and develop PE-VDN. In the first technique, we design a distributed computation scheme that eliminates Vanilla VDN's dependency on sharing environment interaction data. Then, we utilize a privacy-preserving multi-party computation protocol to guarantee that the data flows of the distributed computation scheme do not pose new privacy risks. Finally, we enforce differential privacy to preempt inference threats against the agents' training data—past environment interactions—when they take actions based on their neural network predictions. We implement PE-VDN in StarCraft Multi-Agent Competition (SMAC) and show that it achieves $80\%$ of Vanilla VDN's win rate while maintaining differential privacy levels that provide meaningful privacy guarantees. The results demonstrate that PE-VDN can safeguard the confidentiality of agents' environment interaction data without sacrificing multi-agent coordination. 
\end{abstract}

\input{introduction}

\input{prelims}
\input{related_work}

\input{privacy_engineering}

\input{empirical_results}

\section{CONCLUSIONS \& FUTURE WORK}
In this work, we proposed the PE-VDN algorithm which incorporates three privacy-engineering techniques to protect the confidentiality of the agents' environment interaction data in Vanilla VDN. In particular, we re-engineered the data flows of Vanilla VDN to achieve the following: decentralized training without compromising multi-agent coordination, privacy-preserving inter-agent communication and computation, and differentially private neural network training. 

For future work, we aim to develop similar decentralization and privacy-preserving communication schemes for more sophisticated centralized training and decentralized execution algorithms such as QMIX \cite{rashid2018qmix} and QTRAN \cite{son2019qtran}. These algorithms improve upon the performance of the VDN algorithm by relaxing the additivity assumption in \eqref{eq: additivity assumption}. In particular, the QTRAN algorithm only requires decentralizability as stated in \eqref{eq: decentralizability}.

\bibliographystyle{ieeetr}
\bibliography{refs.bib}


\end{document}

%% file: introduction.tex
\section{INTRODUCTION}
Cooperative multi-agent reinforcement learning (Co-MARL) is a machine learning problem in which multiple agents work together to optimize performance in a common task. The agents interact with an environment that rewards their actions as a team and must learn a decision-making scheme that maximizes the team's long-term rewards through trial and error \cite{yang2020overview}. Co-MARL algorithms can extend the capabilities of single-agent reinforcement learning algorithms to complete complex tasks that involve multiple agents; for instance, in self-driving vehicles where reinforcement learning is a popular approach for autonomous driving \cite{aradi2020survey}, Co-MARL may enable a fleet of autonomous vehicles to cooperate and reduce traffic congestion \cite{fax2004information}.

We study modeling multi-agent coordination in Co-MARL systems with privacy in mind. Effective coordination often requires inter-agent communication and data sharing. However, sharing data may have privacy ramifications in situations where agents represent privacy-sensitive entities or handle privacy-sensitive information. For example, sharing a self-driving vehicle's environment interaction data may reveal commuting patterns and sensitive locations such as home, places of worship, and nightlife activities \cite{li2021privacy}. 

We assume that any data sharing that reveals the agents' interactions with the environment violates privacy. The presence of privacy-sensitive information in the agents' environment interaction data could complicate Co-MARL algorithms that rely on sharing them. Centralized training algorithms \cite{sharma2021survey} are major examples in which a central node consolidates the environment interaction data of all agents and trains a central controller that determines the team's actions. The central controller accounts for the dynamics of how the training of one agent affects others and effectively models multi-agent coordination. By requiring the agents to share their environment interaction data, centralized training methods expose the agents' sensitive information to various privacy risks such as data breaches and unauthorized access. Accordingly, we raise the following research question:
\begin{quote}
    Is it possible to redesign the data flows of centralized training algorithms to safeguard the confidentiality of the agents' environment interaction data without losing multi-agent coordination?
\end{quote}

We show that it is indeed possible to design such an algorithm and propose Privacy-Engineered Value Decomposition Networks (PE-VDN). Instead of relying on a central node that consolidates the agents' environment interaction data, PE-VDN establishes multi-agent coordination by creating peer-to-peer communication channels. Additionally, the algorithm incorporates additional privacy-enhancing techniques to ensure that its established information flows do not undermine the confidentiality of the agents' privacy-sensitive environment interaction data.

We develop PE-VDN based on the Value Decomposition Networks (VDN) algorithm in \cite{sunehag2017value}—referred to as Vanilla VDN, hereafter. Vanilla VDN is a centralized training and decentralized execution (CTDE) Co-MARL algorithm that uses a specially structured function approximator to estimate the team's action values and compute optimal actions. The function approximator consists of multiple branches of neural networks, each of which is designated for one agent. These neural networks use their designated agent's environment interaction data as input, and the summation of their outputs estimates the team's action value. The coupling of the neural networks during training takes multi-agent coordination into account. Once training concludes, the algorithm distributes the final neural network branches back to their designated agents so that they can use them to make their own decisions at runtime. This decentralized execution feature is well-suited to our privacy goals because it eliminates the need for sharing any data during execution. 

We use the Vanilla VDN algorithm as PE-VDN's starting point and incorporate three privacy-engineering techniques to modify data flows and satisfy our privacy goals.
First, we decentralize the vanilla algorithm's training. We demonstrate that the gradients that Vanilla VDN computes for centralized training are only coupled by a summation term that aggregates the output of all neural network branches in the action-value function approximator. We propose an equivalent distributed computation scheme for the gradients that enables the agents to locally maintain and optimize their dedicated neural network branches themselves. The resulting decentralized training algorithm computes the same gradients as Vanilla VDN and only requires the agents to share their neural network outputs with each other; hence, it mitigates the privacy risks of sharing environment interaction data.

In the second privacy measure, we integrate a privacy-preserving multi-party computation protocol with the distributed computation of the gradients. The protocol enables the agents to compute the summation term that couples their gradients while hiding the value of their neural network outputs from each other. Since the agents only need the summation term for decentralized training, revealing their neural network outputs for computing the summation unnecessarily exposes information that correlates with their sensitive environment interaction data. The protocol obfuscates the neural network outputs with correlated random numbers that act as encryption keys, then splits the resulting values into encrypted peer-to-peer messages. The protocol guarantees that as long as no party gains access to all of the peer-to-peer messages that an agent emits, the agents' neural network outputs remain a secret.

The third and last privacy-protection layer that we design for PE-VDN protects the agents' environment interaction data against indirect inference threats. The agents choose their actions based on their internal neural network's predictions and unintentional information leaks can occur when neural networks release their predictions. For example, so-called membership and attribute inference attacks, and inversion attacks are all known to make accurate inferences about the training data of a neural network by mere input and output observations \cite{zhang2022survey}. In PE-VDN, the neural networks that determine the agents' actions are trained with environment interaction data. Our third privacy-engineering technique preempts inference threats against the agents' environment interaction data.

The DP-SGD algorithm \cite{abadi2016deep} is a privacy-preserving training algorithm that can disrupt inference attacks against the training dataset of neural networks. DP-SGD enforces differential privacy during training and helps protect the confidentiality of the training data when external entities interact with the trained models \cite{ha2019differential}. Intuitively, DP-SGD guarantees that two training datasets that differ in a single record produce approximately statistically indistinguishable neural network parameters. If the agents train their internal neural networks with DP-SGD, then the plausible-deniability guarantees that DP-SGD provides can reduce the risks of indirect information leakage about the agents' past environment interaction data.

The DP-SGD algorithm can be seamlessly integrated with our last two privacy-engineering techniques; however, its differential privacy analysis does not readily apply to deep reinforcement learning algorithms. DP-SGD was originally developed for supervised machine learning algorithms with static training datasets, whereas in deep reinforcement learning, a stream of environment interaction data continuously enters a replay buffer and is used for training. We provide a theoretical analysis that leverages DP-SGD's Moments Accountant method \cite{abadi2016deep} and Maximum-Overlap Parallel Composition \cite{smith2021making} to compute the differential privacy level of DP-SGD when applied to deep reinforcement learning. 

We implement a Python library for PE-VDN and test it in the StarCraft Multi-Agent Competition (SMAC) suite \cite{samvelyan19smac}. In the numerical results, we dissect the different privacy-engineering components and study the trade-offs between privacy, precision, and performance. Our results show that within the acceptable differential privacy range of 0 and 10 \cite{ponomareva2023dp}, the agents can achieve $80\%$ of Vanilla VDN's win rate.

%% file: prelims.tex
\section{PRELIMINARIES}
In this section, we cover some technical background on Co-MARL's problem formulation and review Vanilla VDN. Then, we describe our privacy objectives for Co-MARL in the format of a technical privacy policy. 

\subsection{Dec-POMDP}
Co-MARL algorithms typically model multi-agent cooperation using decentralized partially observable Markov decision processes (Dec-POMDPs) \cite{guicheng2022review}.
A Dec-POMDP $\mathcal{G}$ is a tuple $\mathcal{G}=\langle \mathcal{N, S, A, O, Z, P, R}, \gamma \rangle$ where $\mathcal{N}$ is the set of agents and $N=|\mathcal{N}|$; $\mathcal{S}$, $\mathcal{A}$, and $\mathcal{O}$ are the sets of all possible states, actions, and observations, respectively; $\mathcal{Z}:\mathcal{S}\mapsto\mathcal{O}$ is an observation function; $\mathcal{P}: \mathcal{S}^N\times\mathcal{A}^N\times\mathcal{S}^N\mapsto [0,1]$ is the environment's transition probabilities; $\mathcal{R}:\mathcal{S}^N\times\mathcal{A}^N \times \mathcal{S}^N\mapsto\mathbb{R}$ is the reward function; and $\gamma\in [0,1)$ is the discount factor.

A team policy, denoted $\boldsymbol{\pi}:=(\pi_i)_{i\in\mathcal{N}}$, determines the actions that each of the agents must take at every environment observation.
Similar to the notation used for the team policy, we use bold symbols to denote team actions and observations.
In POMDPs, the policy typically incorporates a \textit{history} of past environment observations and actions.
Let $\bs\tau_t =\langle (\bs o_1,\bs a_1), \dots, (\bs o_{t-1}, \bs a_{t-1}), \bs o_t \rangle$ denote the team's history at time $t$ where, for all $k\le t$, $\bs a_k\in \mathcal{A}^N$ and $\bs o\in\mathcal{O}^N$.
Then, $\bs\pi(\bs a_t\mid \bs \tau_t)$ is the probability that the team takes action $\bs a_t$ when its history is $\bs \tau_t$.

Given a team policy $\boldsymbol{\pi}$, a value function $V^{\boldsymbol{\pi}}$ evaluates the expected total reward that the policy accumulates, i.e.,
\begin{equation}
V^{\boldsymbol{\pi}}(\bs s) =
\mathbb{E}\left[\sum_{t=1}^\infty \gamma^{t-1} \mathcal{R}( \bs s_t,\boldsymbol{a}_t,  \bs s_{t+1}) 
\mid \bs s_1 =  \bs s\right],
\end{equation}
where $\boldsymbol{a}_t \sim \boldsymbol{\pi}$ and $\bs s_{t+1}\sim \mathcal{P}(\bs s_{t+1}\mid \bs s_t, \boldsymbol{a_t})$.
An action-value function $Q^{\boldsymbol{\pi}}$ is similarly defined as 
\begin{equation}
    Q^{\boldsymbol{\pi}}(\bs s,\boldsymbol{a}) = \mathbb{E}\left[\mathcal{R}(\bs s, \boldsymbol{a}, \bs s') + \gamma V^{\boldsymbol{\pi}}\left(\bs s'\right)\mid \bs s'\sim \mathcal{P}(\bs s'\mid \bs s, \boldsymbol{a})\right].
\end{equation}
The goal of solving a Dec-POMDP is to find an optimal action-value function $Q^*$ and a corresponding optimal policy $\boldsymbol{\pi}^*$ such that $Q^{\boldsymbol{\pi}^*}(\bs s,\boldsymbol{a}) = Q^*(\bs s,\boldsymbol{a}) = \sup_{\boldsymbol{\pi}} Q^{\boldsymbol{\pi}}(\bs s,\boldsymbol{a})$.
In Co-MARL, the goal is to find an optimal policy without knowing the underlying transition probabilities.

\subsection{Deep Q-Learning}

Deep Q-Learning (DQN) \cite{mnih2015human} is a variant of tabular Q-learning that uses deep neural networks to approximate the Q-function. By using neural networks, DQN can handle high-dimensional and continuous state spaces. DQN uses a \textit{replay buffer} to train the neural network that approximates the Q-function. The replay buffer is a fixed-length database of the agent's past environment interaction data in the form of $\langle s_t, a_t, s_{t+1}, \mathcal{R}(s_t,a_t,s_{t+1})\rangle$, or in short $\langle s, a, s', r\rangle$. The replay buffer is typically filled as follows: the most recent data replaces the oldest entry. However, other methods that use specific heuristics to identify the entry that must be replaced with incoming data exist as well \cite{dao2019relevant}.

With $E$ denoting a minibatch of the replay buffer, DQN minimizes the Bellman error loss function defined as follows:
\begin{equation} \label{eq: single DQN Loss Function}
    \ell(\theta) = \!\!\!\!\!\!\! \sum_{\langle s, a, s', r\rangle\in E} \!\!\! \left(r + \gamma \max_{a'}Q^{\policy}\left(s', a';\theta^-\right) - Q^{\policy}(s,a;\theta)\right)^2,
\end{equation}
where $\theta^-$ is called the target parameters. These parameters are copied from $\theta$ periodically to stabilize training.

\subsection{Multi-Agent DQN}
Both tabular Q-learning and DQN can be used to solve Co-MARL problems. Since the agents in Co-MARL are rewarded together as a team, a single Q-function can represent the action values of the entire team, i.e., the team can be treated as a single agent with a multi-dimensional action. Such bundling of the agents reduces Co-MARL to single-agent reinforcement learning and is the main basis of centralized training methods. Bundling agents requires a consolidated replay buffer to support DQN. In the consolidated replay buffer, all parameters except the rewards are replaced with their team versions—team states and team actions and thus $\langle \bs{s}, \bs{a}, \bs{s'}, r\rangle$ is used instead of $\langle s, a, s', r\rangle$. We also replace all state values $s$ with histories $\tau$ to support partial observability. The resulting loss function is
\begin{equation} \label{eq: DQN Loss Function}
    \sum_{\langle \bs{\tau}, \bs{a}, \bs{\tau'}, r\rangle\in E} \!\!\! \left(r + \gamma \max_{\bs{a'}}Q^{\jpolicy}\left(\bs{\tau'}, \bs{a'};\theta^-\right) - Q^{\policy}(\bs{\tau},\bs{a};\theta)\right)^2.
\end{equation}

Once training concludes, the trained Q-function for the team supports the decision-making of all agents. In this case, the agents must aggregate their environment observations to execute the learned policy too. Therefore, this algorithm is an instance of centralized training and centralized execution.

Alternatively, each agent may attribute the team rewards to itself and use DQN to train a policy independently. This approach is often called independent Q-learning. Independent training allows the agents to execute their policy without having to share their environment observations; however, as opposed to centralized training, independent training does not take multi-agent cooperation into account but it often scales better with the number of agents \cite{yang2020overview}.

In CTDE methods, the agents use centralized training to learn the team's optimal Q-function but then decompose it into a set of local Q-functions that allow for decentralized execution. The agents in CTDE methods choose the action that maximizes their local Q-function; therefore, similar to independent Q-learning, the agents need not share data to execute the team's policy. The difference between these local Q-functions and those obtained via independent training, however, is that the former takes multi-agent coordination into account by design during training. 

CTDE methods typically assume that the agents' individually optimal actions amount to the optimal action for the team. This so-called decentralizability assumption justifies decomposing the team's Q-function into local Q-functions and supports decentralized execution. The formal definition of decentralizability is as follows:
\begin{definition}\label{def: DCT}
    A reinforcement learning task is decentralizable if there exists a collection of local action-value functions $\{Q_i\}_{i \in \mathcal{N}}$ such that, for all team histories $\boldsymbol{\tau}$, team actions $\boldsymbol{a}$, and agents $i\in\mathcal{N}$,
    \begin{equation} \label{eq: decentralizability}
        \left(\mathrm{arg} \max_{\boldsymbol{a}} Q^{\boldsymbol{\pi}}(\boldsymbol{\tau},\boldsymbol{a})\right)_i = \mathrm{arg}\max_{a_i} Q_i(\tau_i,a_i).
    \end{equation}
\end{definition}

\subsection{Vanilla VDN}
The Vanilla VDN algorithm leverages decentralizability by approximating the team's central Q-function with the summation of some local $Q_i$ functions. That is, the algorithm assumes that there exist $\localQ$ such that, for all joint policies $\jpolicy$, histories $\jtraj$, and actions $\jaction$, 
\begin{equation}\label{eq: additivity assumption}
Q^{\jpolicy}(\boldsymbol{\tau},\boldsymbol{a}) = \sum_{i\in\mathcal{N}} Q_i(\tau_i,a_i).
\end{equation} 
This additivity assumption implies decentralizability in \eqref{eq: decentralizability}; however, not all decentralizable tasks satisfy additivity.

In the centralized training phase of Vanilla VDN, the agents must send their environment interaction data to a central node to create a consolidated replay buffer and train the team's Q-function. The optimizer first draws a minibatch of the consolidated replay buffer, denoted $E$. Let, $e=\langle \bs{\tau}, \bs{a}, \bs{\tau'}, r\rangle \in E$ be an element of the minibatch, where $\bs{\tau} = \left( \tau_i \right)_{i\in\mathcal{N}}$, $\bs{a} = \left( a_i \right)_{i\in\mathcal{N}}$, and $\bs{\tau'} = \left( \tau'_i \right)_{i\in\mathcal{N}}$ denote the team's history, action, and next-step history, respectively. With $\theta_i$ denoting the parameters of agent $i$'s dedicated portion of the Q-function, Vanilla VDN's objective function is as follows: 
\begin{multline} \label{eq: VDN Loss Function}
    \ell_{VDN}(e ;\{\theta_i\}_{i\in\mathcal{N}}) = \\
     \left( r +  \sum_{i\in\mathcal{N}} \left(\gamma \max_{a'} Q_i\left(\tau'_i,a';\theta^-_i\right) -  Q_i\left(\tau_i,a_i; \theta_i\right) \right)\right)^2.
\end{multline}

Vanilla VDN uses backpropagation to compute the gradient of the objective function with respect to each $\theta_i$ and uses gradient descent methods to update the Q-function's weights and biases. That is, for all neural network branches $\{\theta_i\}_{i\in\mathcal{N}}$, the optimizer performs the update
\begin{equation} \label{eq: gradient update rule}
    \theta_i \leftarrow \theta_i - \alpha \sum_{e\in E}  \frac{\partial \ell_\text{VDN}(e; \theta_1,\dots, \theta_N)}{\partial \theta_i}, 
\end{equation}
where $\alpha$ is the learning rate. Once centralized training concludes, the optimizer distributes the agents' designated portions—namely $Q_i\left(\cdot,\cdot;\theta_i\right)$ for each $i\in\mathcal{N}$—back to the agents to support decentralized execution.


\subsection{Technical Privacy Policy}
According to the Contextual Integrity Theorem \cite{nissenbaum2019contextual}, privacy is a socially created need and a contextual concept. The theorem explains that \textit{inappropriate information flows} lead to privacy concerns in society. What constitutes inappropriate information flows according to the theorem depends on five parameters: the subject of the information flow, its senders and recipients, information types, and transmission principles such as obtaining consent. Besides privacy concerns, the theorem contends that the interests of the affected parties, ethical and political values, and contextual functions, purposes, and values ultimately determine the ethical legitimacy of information flows in society. 

Based on Contextual Integrity, algorithms do not have innate privacy issues because they can be deployed in various contexts and may or may not establish inappropriate information flows. As a result, we approach privacy for Co-MARL with a scenario-based privacy analysis methodology. That is, we assume that certain privacy concerns exist and develop algorithmic solutions to address them. 

We assume that the agents' environment interaction data in Co-MARL contain privacy-sensitive information such that any information flow that carries this information type is inappropriate. The assumption implies that the existence of a central node that consolidates replay buffers in centralized training algorithms violates privacy. Moreover, any information flow that does not necessarily carry environment interaction data but exposes it to indirect inference reasoning violates privacy as well. 

We formalize our privacy objectives in the format of a technical privacy policy. This policy is not the same as those displayed on websites, but, similar to R. Anderson's security policy for security engineering \cite{anderson2020security}, it is rather a short and verifiable statement of the specifications that our privacy assumptions imply for Co-MARL. The technical privacy policy that we consider is as follows:
\begin{itemize}
    \item Sharing environment interaction data is forbidden.
    \item Under mathematical guarantees that ensure the confidentiality of the sender agent's environment interaction data and that of other agents, the agents may share other types of data for higher team rewards.
\end{itemize}

%% file: related_work.tex
\section{RELATED WORK}

Our main research objective is to design a Co-MARL algorithm that is subject to a technical privacy policy governing the handling of the agents' environment interaction data. In this section, we briefly review the key research themes that are relevant to our research objective.

The survey in \cite{ma2022trusted} and its references to previous related surveys provide an overview of privacy and security for multi-agent machine learning. These surveys point to numerous privacy threats against distributed learning systems and are closely related to the technical privacy policy that we consider. Furthermore, these surveys refer to federated learning, secure multi-party computation, and differential privacy as three prominent defensive mechanisms for general machine learning tasks. We incorporate all three mechanisms in the design of PE-VDN.

Next, we review existing works in designing privacy-aware deep reinforcement learning agents. The single-agent deep reinforcement learning algorithm in \cite{wang2019privacy} is a differentially private algorithm that protects the confidentiality of the agent's rewards. The work in \cite{vietri2020private} develops a single-agent tabular reinforcement learning algorithm that satisfies joint differential privacy—a relaxation of conventional differential privacy. As opposed to the first work, we consider the multi-agent setup, protect the agents' environment interaction data which includes rewards, and instead of developing a customized differential privacy mechanism that perturbs the objective function, we use verified and open-source libraries. Compared with the second algorithm, we consider a multi-agent setup, take continuous observations into account, and do not use differential privacy relaxations. The single-agent algorithm in \cite{zhou2022differentially} generalizes enforcing joint differential privacy to continuous state and action spaces; however, it is restricted to linear function approximators as opposed to this work's use of deep neural networks.

The decentralized multi-agent deep reinforcement learning algorithms in \cite{qu2019value} and \cite{ye2021scalable} closely relate to the technical privacy policy that we consider and both of the algorithms satisfy the policy's first requirement. However, the algorithms do not consider team rewards and assume that the agents are rewarded individually. Moreover, the algorithms do not address the privacy risks associated with the inter-agent communication frameworks that they propose.

%% file: privacy_engineering.tex
\section{Privacy-Engineering Vanilla VDN}
We develop three privacy-engineering techniques to redesign Vanilla VDN's information flows and satisfy the privacy requirements that we expressed in the technical privacy policy. In this section, we describe each of the three technical privacy-enhancing techniques that we use to develop PE-VDN. 

\subsection{Decentralized Training} \label{PE-VDN I}
The Vanilla VDN algorithm's update rule in \eqref{eq: gradient update rule} requires the loss function's gradients with respect to the parameters of all neural network branches within the function approximator. Recall that, in Vanilla VDN, these branches are dedicated to specific agents for decentralized execution. Evaluating the gradient of the loss function in \eqref{eq: VDN Loss Function}, for every branch $\theta_i$ and minibatch sample $e$, we can write
\begin{multline} \label{eq: VDN gradient}
    \frac{\partial \ell_\text{VDN}(e; \theta_1,\dots, \theta_N)}{\partial \theta_i} = \\
     -2  \underbrace{\left( r +  \sum_{i\in\mathcal{N}} \left(\gamma\max_{a'} Q_i\left(\tau'_i,a';\theta^-_i\right) -  Q_i\left(\tau_i,a_i; \theta_i\right) \right)\right)}_{A} \\
    \cdot\underbrace{\frac{\partial Q_i(\tau_i, a_i; \theta_i)}{\partial \theta_i}}_{B}.
\end{multline}

From the gradient expression and its separation into terms A and B, we can observe that the A term is the only factor that couples the gradients of the branches and is a function of all of the agents' environment interaction data. The B term, however, is the gradient of the $i^{\text{th}}$ branch, and computing it only requires the parameters of that branch and its dedicated agent's environment interaction data. 

In PE-VDN, the agents maintain and train their dedicated branch of the VDN themselves. They compute their local gradients by cooperating with other agents to compute the A term in \eqref{eq: VDN gradient} together, which is the coupling term that accounts for multi-agent coordination. We now show how the agents can cooperate to compute the A term without sharing their environment interaction data. 

We require every agent $i$ to compute the message
\begin{equation} \label{eq: q_i messages}
    m_i = \gamma \max_{a'} Q_i\left(\tau'_i,a';\theta^-_i\right) -  Q_i\left(\tau_i,a_i; \theta_i\right)
\end{equation}
and share it with all other agents. Each $m_i$ value can be computed using the $i^{\text{th}}$ agent's environment interaction data and the current parameters of its local $ Q_i$ function. If every agent broadcasts its message $m_i$ to the other agents and locally computes its B term, it can then update its $Q_i$ function just as Vanilla VDN would have updated that agent's dedicated branch. The resulting distributed computation scheme eliminates the need for sharing environment interaction data with a central node and satisfies the first requirement of our technical privacy policy.

\subsection{Privacy-Preserving Multi-Party Summation} \label{PE-VDN II}
Although the summation in the A term can be easily computed by requiring every agent to broadcast their $m_i$ messages, doing so may lead to unintentional information leakage about the agents' environment interaction data. The $m_i$ values are functions of their corresponding agent's privacy-sensitive environment interaction data and correlate with them. We now show how the agents can compute the desired summation in term A in \eqref{eq: VDN gradient} while hiding their $m_i$ values from one another. 

We use a privacy-preserving multi-party computation technique called secret sharing \cite{10.1145/359168.359176} to compute the A term while hiding the underlying $m_i$ values.
Secret sharing refers to the process of dividing a secret into $n$ pieces such that the observation of the pieces reveals no information about the underlying secret unless a sufficient number of them are available. Precisely, a $(k, n)$-secret sharing of $s$ guarantees that observers with access to up to $k-1$ shares can learn no information about $s$. 

We use additive secret sharing which is an efficient $(n,n)$-secret sharing scheme \cite{xiong2020efficient}. In particular, let $s$ be the secret value over a finite filed $\mathbb{Z}_p := \{0, 1, \dots, p-1\}$. Then, additive secret sharing splits the secret $s\in\mathbb{Z}_p$ into $n$ shares 
$S(s)=(r_1, r_2, \dots, r_n)$ such that $r_1,\dots,r_{n-1}$ are chosen uniformly at random from $\mathbb{Z}_p$, and $r_n = p - \left(\sum_{i=1}^{n-1} r_i \mod p\right) + s$.
Additive secret sharing ensures that a complete set of $n$ shares accurately reconstruct the secret via $S^{-1}(r_1,\dots,r_n) := \left(\sum_{i=1}^{n} r_i\right) \mod p$. Otherwise, for any given set of shares with less than $n$ shares, every element of $\mathbb{Z}_p$ is equally likely to generate those shares. Additive secret sharing is \textit{additively homomorphic}, i.e., for secrets $s_1, \dots, s_m$, all in $\mathbb{Z}_p$, we have that
\begin{equation} \label{eq: pp summation}
    \sum_{i=1}^m s_i = S^{-1}\left( \sum_{i=1}^m S(s_i) \right),
\end{equation}
where the right-hand side of \eqref{eq: pp summation} is the element-wise summation of the secret shares.

The additive homomorphism of additive secret sharing ensures the accuracy of the following three-step privacy-preserving $n$-party summation protocol: first, each party invokes additive secret sharing on its secret value and sends each of the other agents a piece of the shares. Then, the agents locally compute the sum of the shares that they receive from other agents and broadcast the results. Finally, the agents recover the summation of their secret values by computing the sum of the broadcast values.

Returning to the problem of computing the A term in the gradient expression in \eqref{eq: VDN gradient}, the privacy-preserving summation protocol based on additive secret sharing can be deployed with one intermediate step. Additive secret sharing applies to finite-field integers and we need to encode the $m_i$ values in that format. We use the encoding
\begin{equation} \label{eq: encoding}
    \mathrm{enc}(x) = \mathrm{int} \left(10^{\texttt{PRECISION}} \cdot x\right) \mod p,
\end{equation}
in which $\texttt{PRECISION}$ denotes the number of decimal places that the encoding preserves. To reverse the encoding back to floating point, we use the decoding function
\begin{equation} \label{eq: decoding}
    \mathrm{dec}(x) = \begin{cases}x/10^{\texttt{PRECISION}} \quad &\text{if } x\le p/2, \\ (x-p)/10^{\texttt{PRECISION}} \quad &\text{if } x>p/2.  \end{cases}
\end{equation}

To summarize, in order to compute the summation in the A term, the agents send the secret shares $S(\mathrm{enc}(m_i))$ to one another via peer-to-peer messaging. Then, each agent computes the summation of its received shares and broadcasts the result. Finally, each agent reconstructs a copy of the A term with the summation of the broadcast values. Integrating additive secret sharing with the distributed computation of the gradients ensures that inter-agent communications in the decentralized training of PE-VDN do not undermine the confidentiality of the agents' environment interaction data.

\subsection{Training with Differential Privacy} \label{PE-VDN III}

When the agents choose actions based on their neural network's action-value predictions, they may unintentionally leak information about their environment interaction data as well. It has been observed that the predictions of neural networks could mirror specific relationships in the training dataset that were not intended to be exposed \cite{zhang2021survey}. For example, text-generative neural networks may complete certain input prompts with memorized phrases within their training datasets \cite{jayaraman2022active}. Similarly, the actions of deep reinforcement learning agents—including independent training algorithms in Co-MARL—may reveal certain sensitive characteristics of their training experience as well \cite{pan2019you}. 


A machine learning algorithm that trains a neural network must map a training dataset to a set of parameters that determine the neural network's weights and biases. Under fixed hyperparameters and pseudo-randomization seeds, training algorithms are deterministic mappings. These deterministic mappings may produce outputs that external observers can accurately associate with certain training data. Privacy-enhancing techniques based on differential privacy can disrupt inference privacy threats against the training datasets \cite{zhang2021survey}. Differentially private training algorithms can provably generate model parameters that are approximately statistically indistinguishable from those trained with datasets that differ in only one element. This guarantee establishes a plausible deniability argument against the accuracy of privacy threats that base their attacks on the learned parameters of a neural network or its predictions. 

A formal definition of $(\epsilon, \delta)$-differential privacy is as follows: let $(\Omega, \mathcal{F}, \Pr)$ be a probability space and $f: \Omega \times \mathcal{D} \mapsto \mathcal{R}$ be a stochastic mapping that maps a dataset $D\in\mathcal{D}$ to its image under $f$. The mapping $f$ satisfies $(\epsilon, \delta)$-differential privacy if, for all datasets $D\in\mathcal{D}$ and $D'\in\mathcal{D}$, whose elements are identical except for one, and all subsets $S\subseteq \mathcal{R}$,
\begin{equation} \label{eq: dp}
    \Pr[f(D) \in S] \leq \exp({\epsilon}) \cdot \Pr[f(D') \in S] + \delta.
\end{equation}


The DP-SGD algorithm \cite{abadi2016deep} is a differentially private supervised learning algorithm for neural networks and enforces differential privacy by repeatedly injecting calibrated Gaussian noise into the gradients. 
The algorithm's so-called Moments Accountant subroutine tracks the differential privacy level—$\epsilon$ and $\delta$—of the composition of the iterations throughout the training.

As opposed to typical gradient descent algorithms that draw fixed-length minibatches, DP-SGD uses Poisson sampling, which refers to a sampling method that chooses each of the minibatch elements with a fixed probability. Moreover, DP-SGD clips the gradients to a fixed threshold $C$. That is, with $g_x$ denoting the gradient for sample $x$, DP-SGD uses the mapping $\mathrm{Clip}_C(g_x) := g_x \cdot \max\left(1, {\|g_x\|_2}/{C}\right)^{-1}$.

\input{algorithm}

We integrate DP-SGD with our earlier decentralization and the privacy-preserving multi-party summation protocol contributions as follows: instead of performing Poisson sampling on a static labeled dataset as in supervised learning, we perform Poisson sampling on the stream of environment interaction data that gets loaded onto the agents' replay buffers. The integration of the DP-SGD concludes the design of PE-VDN and Algorithm \ref{alg: contribution} describes its pseudocode.

The application of DP-SGD to the contents of the agents' replay buffers makes the existing differential privacy analysis of DP-SGD ill-suited to PE-VDN. Specifically, the differential privacy analysis of the DP-SGD algorithm via the Moments Accountant method \cite{abadi2016deep} only applies to the contents of the replay buffer for a single iteration, not the agents' entire collection of environment interaction data. As the last remaining puzzle piece, we now state a theorem that computes the differential privacy level of PE-VDN.

\begin{theorem} \label{thm}
Fix a Poisson sampling rate $q$, noise variance $\sigma^2$, and $\delta$. Let $(\epsilon(T),\delta)$ be the differential privacy level that the Moments Accountant method computes for DP-SGD under $q$, $\sigma^2$, and $\delta$ after $T$ iterations. Then, with the same parameters, PE-VDN is $(\Tilde{\epsilon}, \Tilde{\delta})-$differentially private with $\Tilde{\epsilon} = \texttt{buffer\_throughput}^{-1} \cdot \epsilon(\texttt{buffer\_size})$, and $\Tilde{\delta} = \texttt{buffer\_throughput}^{-1} \cdot \delta$.
\end{theorem}

Due to space limitations, we refer the reader to the extended version of the paper\footnote{See \href{https://drive.google.com/drive/folders/1fa9ptiWYrAmiBe_a8hadmluJvD34L1Cg?usp=share_link}{here}.} for the full proof of the Theorem. Intuitively, the proof leverages the property that once an episode leaves the replay buffer, it no longer remains part of the training. The proof formalizes this property in the context of Maximum Overlap Composition Theorem in \cite{smith2021making} to compute the differential privacy of Algorithm \ref{alg: contribution}.

%% file: algorithm.tex
\SetKwComment{Comment}{/* }{ */}

\begin{algorithm} [!t] \label{alg: contribution}
    \caption{PE-VDN Pseudocode}
\KwIn{Team of Agents, Environment $\mathcal{E}$, Hyperparameters $\mathcal{H}$}
\KwOut{Trained Agents}
 $\texttt{total\_steps} \gets 0$\;
 $q\gets\texttt{expected\_batch\_size} / \texttt{buffer\_size}$\;
\While{$\texttt{total\_steps} < \texttt{max\_steps}$}{
RUN an episode and add the episode length to \texttt{total\_steps} \;
\For{every agent $i$}{
APPEND environment interaction data $\langle \tau_i, a_i, \tau'_i, r\rangle$ to local replay buffer\; 
}
CHOOSE minibatch indices $I$ using Poisson sampling with sample rate $q$\;
BROADCAST the indices to all agents\;
\For{every agent $i$}{ 
COMPUTE $m_i$-values for all entries in $I$ using \eqref{eq: q_i messages} \;
ENCODE $m_i$-values using \eqref{eq: encoding} \;
ENCRYPT the encodings using additive secret sharing\;
STORE the $i^{\text{th}}$ share locally\;
DISTRIBUTE the remaining shares\;
WAIT for all agents to distribute their shares\;
COMPUTE the summation of the received shares and the locally stored share\;
BROADCAST the outcome\;
WAIT for all agents to broadcast theirs\;
DECRYPT the broadcast values by reversing additive secret sharing\;
DECODE the results using \eqref{eq: decoding} to compute term A in \eqref{eq: VDN gradient}\;
COMPUTE local gradients using \eqref{eq: VDN gradient}\;
AVERAGE gradients along the episode for every minibatch sample\; 
CLIP gradients with threshold $C$\;
AGGREGATE clipped gradients across minibatch samples\;
ADD zero-mean Gaussian Noise with $\sigma = \texttt{noise\_multiplier} \cdot C$\ to the aggregated gradients\;
UPDATE parameters using SGD with probability $1/\texttt{buffer\_throughput}$\;
}
}
\end{algorithm}

%% file: empirical_results.tex
\section{Numerical Experiments}
Based on the pseudocode in Algorithm \ref{alg: contribution}, we implement a Python library for PE-VDN\footnote{All codes and instructions are provided \href{https://github.com/parhamgohari/PrivacyCoMARL.git}{here}.}.
 We use the StarCraft Multi-Agent Competition (SMAC) suite \cite{samvelyan19smac}, a machine learning application programming interface (API) for using Co-MARL to learn how to play StarCraft II. Our experiments take place in SMAC's \texttt{3m} environment in which three agents must cooperate to kill three pre-trained enemies.

In the experiments, we dissect the different components of PE-VDN to empirically evaluate their effects on team rewards. First, we consider the decentralization of Vanilla VDN's training. We do not expect that decentralization will affect performance because the distributed computation scheme in Section \ref{PE-VDN I} (PE-VDN A) computes the same gradients as Vanilla VDN. We also do not anticipate any significant performance drop due to the privacy-preserving multi-party summation protocol (PE-VDN B) because the only source of inaccuracy that it introduces is the encoding function's quantization error. 

\begin{figure*}[t]
    \centering
    \input{plots/centralized_vs_decentralized}
    \input{plots/dp}
    \caption{Comparison of the win rate of different algorithms in SMAC's \texttt{3m} environment. The left plot compares win rates of Vanilla VDN, Independent Q-Learning (IQL), PE-VDN with only decentralized training (PE-VDN A), and PE-VDN with decentralized training and privacy-preserving multi-party summation (PE-VDN B) algorithms. The right plot demonstrates the win rate of PE-VDN with differential privacy (PE-VDN C). The solid line represents the anchor model's win rate and the volatile dashed line is the win rate of the running model.}
    \label{fig:1}
\end{figure*}
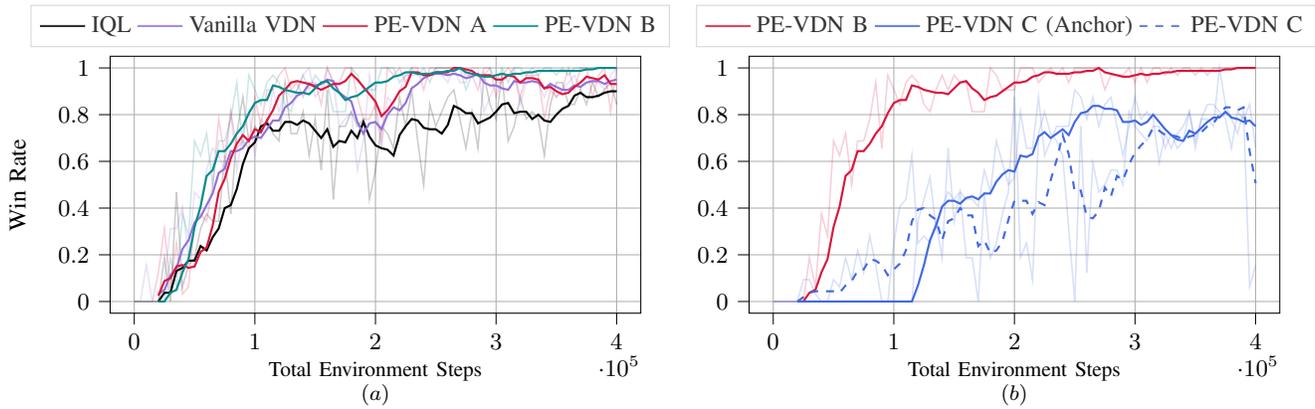

We periodically evaluate the agents' win rate during training and plot the results in Figure \ref{fig:1}. The win rates confirm that Vanilla VDN, PE-VDN A, and PE-VDN B with $\texttt{PRECISION}=5$ all perform similarly. Moreover, recall that the technical privacy policy does not allow any data sharing that does not benefit the agents in terms of the team rewards. The superior performance of Vanilla VDN, PE-VDN A, and PE-VDN B compared with Independent Q-Learning in Figure \ref{fig:1}a indicates that the agents indeed benefit from cooperation. Although, Vanilla VDN violates the first prong of the technical privacy policy.

In the next experiment, we evaluate the effects of enforcing differential privacy via the DP-SGD algorithm. DP-SGD's injection of noise into the gradients may affect performance negatively. In the supervised learning setup where DP-SGD has been primarily studied, it has been reported that the performance of the models trained with DP-SGD is more sensitive to the choice of training hyperparameters than non-differentially private counterparts \cite{ponomareva2023dp}. In particular, hyperparameter tuning may play a key role in model accuracy, training stability, and sample complexity \cite{ponomareva2023dp}. Clipping gradients may add bias, perturbing the gradients may deflect the parameters away from local minima and destabilize training, and achieving meaningful $\epsilon$ and $\delta$ values—$\epsilon<3$ and $\delta < 1/n^{1.1}$, where $n$ is the training dataset's size \cite{ponomareva2023dp}—may require large training datasets.

In our implementations, we use the Opacus \cite{opacus} toolbox which is an open-source library for DP-SGD. We use Opacus' built-in Moments Accountant method to track PE-VDN's differential privacy levels as stated in Theorem \ref{thm}. Without differential privacy, we used the Adam optimizer with a learning rate of $5\cdot 10^{-4}$. In light of the guidelines in \cite{ponomareva2023dp}, we found that using the SGD optimizer with weight decay $0.01$, momentum $0.9$, and a significantly higher learning rate $5\cdot10^{-3}$ performs better for DP-SGD. Moreover, we use an \textit{anchoring technique} to improve the training's stability further. In this technique, once the periodic win-rate evaluation returns rates that are beyond a specific threshold, we save the parameters as an anchor model. Then we repeatedly increase the threshold and penalize the rest of the training with the current parameters' distance from the anchor. With these techniques, Figure \ref{fig:1}b shows the performance of PE-VDN under $(2.90,4.9\cdot10^{-4})$-differential privacy and benchmarks it with PE-VDN B with Poisson sampling.



%% file: plots/centralized_vs_decentralized.tex
\begin{tikzpicture}

\definecolor{crimson}{RGB}{220,20,60}
\definecolor{darkcyan}{RGB}{0,139,139}
\definecolor{darkgray176}{RGB}{176,176,176}
\definecolor{mediumpurple}{RGB}{147,112,219}
\definecolor{royalblue}{RGB}{0,0,0}

\begin{axis}[
legend cell align={left},
legend style={fill opacity=0.8, draw opacity=0.8, text opacity=1, draw=white!80!black, font = \small, at = {(-0.15,1.2)}, anchor = north west},
legend columns = 4,
legend entries = {,,,,IQL, Vanilla VDN, PE-VDN A, PE-VDN B},
tick align=outside,
tick pos=left,
unbounded coords=jump,
x grid style={darkgray176},
xlabel={$\substack{\text{Total Environment Steps}\\(a)}$},
xmajorgrids,
xmin=-20000, xmax=420000,
xtick style={color=black},
y grid style={darkgray176},
ylabel={Win Rate},
ytick = {0, 0.2, 0.4, 0.6, 0.8, 1},
ymajorgrids,
ymin=-0.05, ymax=1.05,
ytick style={color=black},
height = 5cm,
width = \columnwidth,
xlabel style = {font = \large},
ylabel style = {font = \small},
ticklabel style = {font = \small}
]
\addplot [semithick, royalblue, opacity=0.2]
table {%
0 0
5000 0
10000 0
15000 0
20000 0
25000 0.1875
30000 0
35000 0.46875
40000 0.0625
45000 0.15625
50000 0.1875
55000 0.3125
60000 0.375
65000 0.34375
70000 0.34375
75000 0.625
80000 0.375
85000 0.71875
90000 0.875
95000 0.6875
100000 0.75
105000 0.71875
110000 0.78125
115000 0.71875
120000 0.6875
125000 0.9375
130000 0.71875
135000 0.71875
140000 0.8125
145000 0.625
150000 0.78125
155000 0.5625
160000 0.90625
165000 0.4375
170000 0.75
175000 0.75
180000 0.8125
185000 0.75
190000 0.78125
195000 0.4375
200000 0.5625
205000 0.75
210000 0.71875
215000 0.65625
220000 0.875
225000 0.75
230000 0.90625
235000 0.6875
240000 0.5
245000 0.84375
250000 0.75
255000 0.8125
260000 0.875
265000 0.90625
270000 0.78125
275000 0.65625
280000 0.8125
285000 0.65625
290000 0.96875
295000 0.84375
300000 0.78125
305000 0.96875
310000 0.6875
315000 0.75
320000 0.875
325000 0.78125
330000 0.8125
335000 0.78125
340000 0.625
345000 0.84375
350000 0.9375
355000 0.9375
360000 0.84375
365000 0.84375
370000 0.90625
375000 0.84375
380000 0.9375
385000 0.875
390000 0.90625
395000 0.9375
400000 0.84375
};
\addplot [semithick, mediumpurple, opacity=0.2]
table {%
0 0
5000 0
10000 0.15625
15000 0
20000 0
25000 0.09375
30000 0.3125
35000 0.3125
40000 0.40625
45000 0.1875
50000 0.46875
55000 0.4375
60000 0.5625
65000 0.65625
70000 0.625
75000 0.5625
80000 0.8125
85000 0.5625
90000 0.71875
95000 0.8125
100000 0.625
105000 0.78125
110000 0.75
115000 0.90625
120000 0.8125
125000 0.90625
130000 0.875
135000 0.90625
140000 0.96875
145000 1
150000 0.875
155000 0.90625
160000 1
165000 0.9375
170000 0.75
175000 0.78125
180000 0.8125
185000 0.5625
190000 0.6875
195000 0.96875
200000 0.8125
205000 0.65625
210000 0.875
215000 0.84375
220000 0.90625
225000 0.96875
230000 0.96875
235000 0.96875
240000 0.96875
245000 0.96875
250000 1
255000 0.96875
260000 0.9375
265000 1
270000 0.9375
275000 0.9375
280000 1
285000 1
290000 0.8125
295000 0.875
300000 0.9375
305000 0.9375
310000 0.96875
315000 1
320000 0.84375
325000 0.9375
330000 0.96875
335000 0.9375
340000 0.875
345000 0.9375
350000 0.875
355000 0.90625
360000 0.9375
365000 0.96875
370000 0.90625
375000 0.96875
380000 0.9375
385000 0.9375
390000 0.90625
395000 0.96875
400000 1
};
\addplot [semithick, crimson, opacity=0.2]
table {%
0 0
5000 0
10000 0
15000 0
20000 0.125
25000 0.3125
30000 0.0625
35000 0.25
40000 0.03125
45000 0.0625
50000 0.34375
55000 0.375
60000 0.34375
65000 0.5
70000 0.78125
75000 0.625
80000 0.8125
85000 0.65625
90000 0.6875
95000 0.65625
100000 0.875
105000 0.75
110000 0.90625
115000 0.90625
120000 0.96875
125000 1
130000 0.90625
135000 0.9375
140000 0.875
145000 0.9375
150000 0.9375
155000 0.84375
160000 1
165000 0.9375
170000 0.9375
175000 1
180000 1
185000 0.875
190000 0.78125
195000 0.78125
200000 0.84375
205000 0.6875
210000 1
215000 0.96875
220000 1
225000 0.9375
230000 1
235000 1
240000 0.90625
245000 1
250000 1
255000 1
260000 1
265000 1
270000 1
275000 0.96875
280000 0.96875
285000 0.90625
290000 1
295000 1
300000 0.875
305000 0.96875
310000 1
315000 0.96875
320000 0.875
325000 0.96875
330000 0.96875
335000 0.96875
340000 0.78125
345000 0.90625
350000 0.875
355000 0.90625
360000 1
365000 0.90625
370000 1
375000 1
380000 0.875
385000 0.96875
390000 1
395000 0.8125
400000 1
};
\addplot [semithick, darkcyan, opacity=0.2]
table {%
0 0
5000 0
10000 0
15000 0
20000 0
25000 0
30000 0.1875
35000 0.0625
40000 0.375
45000 0.28125
50000 0.6875
55000 0.625
60000 0.71875
65000 0.5
70000 0.6875
75000 0.6875
80000 0.78125
85000 0.9375
90000 0.65625
95000 0.96875
100000 0.90625
105000 0.84375
110000 0.9375
115000 0.96875
120000 0.9375
125000 0.84375
130000 0.8125
135000 0.90625
140000 0.96875
145000 0.90625
150000 0.96875
155000 0.9375
160000 0.9375
165000 0.8125
170000 0.8125
175000 0.8125
180000 1
185000 0.96875
190000 0.90625
195000 0.90625
200000 0.90625
205000 1
210000 1
215000 1
220000 0.9375
225000 0.96875
230000 1
235000 0.96875
240000 1
245000 0.9375
250000 1
255000 1
260000 1
265000 1
270000 1
275000 0.90625
280000 0.96875
285000 0.96875
290000 0.96875
295000 1
300000 0.9375
305000 1
310000 0.9375
315000 1
320000 1
325000 0.96875
330000 1
335000 0.96875
340000 1
345000 1
350000 0.96875
355000 1
360000 0.96875
365000 1
370000 1
375000 1
380000 1
385000 1
390000 1
395000 1
400000 1
};
\addplot [thick, royalblue]
table {%
0 nan
5000 nan
10000 nan
15000 nan
20000 0
25000 0.0375
30000 0.0375
35000 0.13125
40000 0.14375
45000 0.175
50000 0.175
55000 0.2375
60000 0.21875
65000 0.275
70000 0.3125
75000 0.4
80000 0.4125
85000 0.48125
90000 0.5875
95000 0.65625
100000 0.68125
105000 0.75
110000 0.7625
115000 0.73125
120000 0.73125
125000 0.76875
130000 0.76875
135000 0.75625
140000 0.775
145000 0.7625
150000 0.73125
155000 0.7
160000 0.7375
165000 0.6625
170000 0.6875
175000 0.68125
180000 0.73125
185000 0.7
190000 0.76875
195000 0.70625
200000 0.66875
205000 0.65625
210000 0.65
215000 0.625
220000 0.7125
225000 0.75
230000 0.78125
235000 0.775
240000 0.74375
245000 0.7375
250000 0.7375
255000 0.71875
260000 0.75625
265000 0.8375
270000 0.825
275000 0.80625
280000 0.80625
285000 0.7625
290000 0.775
295000 0.7875
300000 0.8125
305000 0.84375
310000 0.85
315000 0.80625
320000 0.8125
325000 0.8125
330000 0.78125
335000 0.8
340000 0.775
345000 0.76875
350000 0.8
355000 0.825
360000 0.8375
365000 0.88125
370000 0.89375
375000 0.875
380000 0.875
385000 0.88125
390000 0.89375
395000 0.9
400000 0.9
};
\addplot [thick, mediumpurple]
table {%
0 nan
5000 nan
10000 nan
15000 nan
20000 0.03125
25000 0.05
30000 0.1125
35000 0.14375
40000 0.225
45000 0.2625
50000 0.3375
55000 0.3625
60000 0.4125
65000 0.4625
70000 0.55
75000 0.56875
80000 0.64375
85000 0.64375
90000 0.65625
95000 0.69375
100000 0.70625
105000 0.7
110000 0.7375
115000 0.775
120000 0.775
125000 0.83125
130000 0.85
135000 0.88125
140000 0.89375
145000 0.93125
150000 0.925
155000 0.93125
160000 0.95
165000 0.94375
170000 0.89375
175000 0.875
180000 0.85625
185000 0.76875
190000 0.71875
195000 0.7625
200000 0.76875
205000 0.7375
210000 0.8
215000 0.83125
220000 0.81875
225000 0.85
230000 0.9125
235000 0.93125
240000 0.95625
245000 0.96875
250000 0.975
255000 0.975
260000 0.96875
265000 0.975
270000 0.96875
275000 0.95625
280000 0.9625
285000 0.975
290000 0.9375
295000 0.925
300000 0.925
305000 0.9125
310000 0.90625
315000 0.94375
320000 0.9375
325000 0.9375
330000 0.94375
335000 0.9375
340000 0.9125
345000 0.93125
350000 0.91875
355000 0.90625
360000 0.90625
365000 0.925
370000 0.91875
375000 0.9375
380000 0.94375
385000 0.94375
390000 0.93125
395000 0.94375
400000 0.95
};
\addplot [thick, crimson]
table {%
0 nan
5000 nan
10000 nan
15000 nan
20000 0.025
25000 0.0875
30000 0.1
35000 0.15
40000 0.15625
45000 0.14375
50000 0.15
55000 0.2125
60000 0.23125
65000 0.325
70000 0.46875
75000 0.525
80000 0.6125
85000 0.675
90000 0.7125
95000 0.6875
100000 0.7375
105000 0.725
110000 0.775
115000 0.81875
120000 0.88125
125000 0.90625
130000 0.9375
135000 0.94375
140000 0.9375
145000 0.93125
150000 0.91875
155000 0.90625
160000 0.91875
165000 0.93125
170000 0.93125
175000 0.94375
180000 0.975
185000 0.95
190000 0.91875
195000 0.8875
200000 0.85625
205000 0.79375
210000 0.81875
215000 0.85625
220000 0.9
225000 0.91875
230000 0.98125
235000 0.98125
240000 0.96875
245000 0.96875
250000 0.98125
255000 0.98125
260000 0.98125
265000 1
270000 1
275000 0.99375
280000 0.9875
285000 0.96875
290000 0.96875
295000 0.96875
300000 0.95
305000 0.95
310000 0.96875
315000 0.9625
320000 0.9375
325000 0.95625
330000 0.95625
335000 0.95
340000 0.9125
345000 0.91875
350000 0.9
355000 0.8875
360000 0.89375
365000 0.91875
370000 0.9375
375000 0.9625
380000 0.95625
385000 0.95
390000 0.96875
395000 0.93125
400000 0.93125
};
\addplot [thick, darkcyan]
table {%
0 nan
5000 nan
10000 nan
15000 nan
20000 0
25000 0
30000 0.0375
35000 0.05
40000 0.125
45000 0.18125
50000 0.31875
55000 0.40625
60000 0.5375
65000 0.5625
70000 0.64375
75000 0.64375
80000 0.675
85000 0.71875
90000 0.75
95000 0.80625
100000 0.85
105000 0.8625
110000 0.8625
115000 0.925
120000 0.91875
125000 0.90625
130000 0.9
135000 0.89375
140000 0.89375
145000 0.8875
150000 0.9125
155000 0.9375
160000 0.94375
165000 0.9125
170000 0.89375
175000 0.8625
180000 0.875
185000 0.88125
190000 0.9
195000 0.91875
200000 0.9375
205000 0.9375
210000 0.94375
215000 0.9625
220000 0.96875
225000 0.98125
230000 0.98125
235000 0.975
240000 0.975
245000 0.975
250000 0.98125
255000 0.98125
260000 0.9875
265000 0.9875
270000 1
275000 0.98125
280000 0.975
285000 0.96875
290000 0.9625
295000 0.9625
300000 0.96875
305000 0.975
310000 0.96875
315000 0.975
320000 0.975
325000 0.98125
330000 0.98125
335000 0.9875
340000 0.9875
345000 0.9875
350000 0.9875
355000 0.9875
360000 0.9875
365000 0.9875
370000 0.9875
375000 0.99375
380000 0.99375
385000 1
390000 1
395000 1
400000 1
};
\end{axis}

\end{tikzpicture}

%% file: plots/dp.tex
\begin{tikzpicture}

\definecolor{crimson}{RGB}{220,20,60}
\definecolor{darkcyan}{RGB}{0,139,139}
\definecolor{darkgray176}{RGB}{176,176,176}
\definecolor{royalblue}{RGB}{65,105,225}

\begin{axis}[
legend cell align={left},
legend columns = 3,
legend style={fill opacity=0.8, draw opacity=0.8, text opacity=1, draw=white!80!black, font = \small, at = {(-0.1,1.2)}, anchor = north west},
legend entries = {,,,PE-VDN B, PE-VDN C (Anchor), PE-VDN C},
tick align=outside,
tick pos=left,
unbounded coords=jump,
x grid style={darkgray176},
xlabel={$\substack{\text{Total Environment Steps}\\(b)}$},
xmajorgrids,
xmin=-20000, xmax=420000,
xtick style={color=black},
y grid style={darkgray176},
ytick = {0, 0.2, 0.4, 0.6, 0.8, 1},
ymajorgrids,
ymin=-0.05, ymax=1.05,
ytick style={color=black},
height = 5cm,
width = \columnwidth,
xlabel style = {font = \large},
ylabel style = {font = \small},
ticklabel style = {font = \small}
]
\addplot [semithick, crimson, opacity=0.2]
table {%
0 0
5000 0
10000 0
15000 0
20000 0
25000 0
30000 0.1875
35000 0.0625
40000 0.375
45000 0.28125
50000 0.6875
55000 0.625
60000 0.71875
65000 0.5
70000 0.6875
75000 0.6875
80000 0.78125
85000 0.9375
90000 0.65625
95000 0.96875
100000 0.90625
105000 0.84375
110000 0.9375
115000 0.96875
120000 0.9375
125000 0.84375
130000 0.8125
135000 0.90625
140000 0.96875
145000 0.90625
150000 0.96875
155000 0.9375
160000 0.9375
165000 0.8125
170000 0.8125
175000 0.8125
180000 1
185000 0.96875
190000 0.90625
195000 0.90625
200000 0.90625
205000 1
210000 1
215000 1
220000 0.9375
225000 0.96875
230000 1
235000 0.96875
240000 1
245000 0.9375
250000 1
255000 1
260000 1
265000 1
270000 1
275000 0.90625
280000 0.96875
285000 0.96875
290000 0.96875
295000 1
300000 0.9375
305000 1
310000 0.9375
315000 1
320000 1
325000 0.96875
330000 1
335000 0.96875
340000 1
345000 1
350000 0.96875
355000 1
360000 0.96875
365000 1
370000 1
375000 1
380000 1
385000 1
390000 1
395000 1
400000 1
};
\addplot [semithick, royalblue, opacity=0.2]
table {%
0 0
5000 0
10000 0
15000 0
20000 0
25000 0
30000 0
35000 0
40000 0
45000 0
50000 0
55000 0
60000 0
65000 0
70000 0
75000 0
80000 0
85000 0
90000 0
95000 0
100000 0
105000 0
110000 0
115000 0
120000 0.34375
125000 0.4375
130000 0.53125
135000 0.3125
140000 0.40625
145000 0.46875
150000 0.4375
155000 0.46875
160000 0.4375
165000 0.4375
170000 0.40625
175000 0.5625
180000 0.46875
185000 0.6875
190000 0.59375
195000 0.5
200000 0.53125
205000 0.8125
210000 0.65625
215000 0.65625
220000 0.875
225000 0.625
230000 0.6875
235000 0.75
240000 0.75
245000 0.75
250000 0.90625
255000 0.875
260000 0.78125
265000 0.875
270000 0.75
275000 0.84375
280000 0.875
285000 0.75
290000 0.625
295000 0.75
300000 0.875
305000 0.78125
310000 0.8125
315000 0.78125
320000 0.59375
325000 0.71875
330000 0.65625
335000 0.71875
340000 0.75
345000 0.8125
350000 0.65625
355000 0.78125
360000 0.84375
365000 0.84375
370000 0.8125
375000 0.78125
380000 0.65625
385000 0.8125
390000 0.78125
395000 0.84375
400000 0.65625
};
\addplot [semithick, royalblue, opacity=0.2]
table {%
0 0
5000 0
10000 0
15000 0
20000 0
25000 0.09375
30000 0.09375
35000 0.03125
40000 0
45000 0
50000 0.09375
55000 0.0625
60000 0.1875
65000 0.09375
70000 0.21875
75000 0.125
80000 0.28125
85000 0.15625
90000 0
95000 0
100000 0.25
105000 0.40625
110000 0.4375
115000 0.625
120000 0.25
125000 0.28125
130000 0.25
135000 0.34375
140000 0.1875
145000 0.65625
150000 0.34375
155000 0.46875
160000 0.1875
165000 0.1875
170000 0
175000 0.46875
180000 0.1875
185000 0.25
190000 0.375
195000 0.46875
200000 0.875
205000 0.1875
210000 0.25
215000 0.09375
220000 0.71875
225000 0.8125
230000 0.71875
235000 0.6875
240000 0.625
245000 0.3125
250000 0
255000 0.75
260000 0.125
265000 0.59375
270000 0.46875
275000 0.5
280000 0.5625
285000 0.5625
290000 0.5
295000 0.8125
300000 0.75
305000 0.71875
310000 0.6875
315000 0.75
320000 0.75
325000 0.65625
330000 0.71875
335000 0.65625
340000 0.6875
345000 0.8125
350000 0.6875
355000 0.78125
360000 0.78125
365000 0.75
370000 1
375000 0.84375
380000 0.78125
385000 0.71875
390000 0.8125
395000 0.0625
400000 0.15625
};
\addplot [thick, crimson]
table {%
0 nan
5000 nan
10000 nan
15000 nan
20000 0
25000 0
30000 0.0375
35000 0.05
40000 0.125
45000 0.18125
50000 0.31875
55000 0.40625
60000 0.5375
65000 0.5625
70000 0.64375
75000 0.64375
80000 0.675
85000 0.71875
90000 0.75
95000 0.80625
100000 0.85
105000 0.8625
110000 0.8625
115000 0.925
120000 0.91875
125000 0.90625
130000 0.9
135000 0.89375
140000 0.89375
145000 0.8875
150000 0.9125
155000 0.9375
160000 0.94375
165000 0.9125
170000 0.89375
175000 0.8625
180000 0.875
185000 0.88125
190000 0.9
195000 0.91875
200000 0.9375
205000 0.9375
210000 0.94375
215000 0.9625
220000 0.96875
225000 0.98125
230000 0.98125
235000 0.975
240000 0.975
245000 0.975
250000 0.98125
255000 0.98125
260000 0.9875
265000 0.9875
270000 1
275000 0.98125
280000 0.975
285000 0.96875
290000 0.9625
295000 0.9625
300000 0.96875
305000 0.975
310000 0.96875
315000 0.975
320000 0.975
325000 0.98125
330000 0.98125
335000 0.9875
340000 0.9875
345000 0.9875
350000 0.9875
355000 0.9875
360000 0.9875
365000 0.9875
370000 0.9875
375000 0.99375
380000 0.99375
385000 1
390000 1
395000 1
400000 1
};
\addplot [thick, royalblue]
table {%
0 nan
5000 nan
10000 nan
15000 nan
20000 0
25000 0
30000 0
35000 0
40000 0
45000 0
50000 0
55000 0
60000 0
65000 0
70000 0
75000 0
80000 0
85000 0
90000 0
95000 0
100000 0
105000 0
110000 0
115000 0
120000 0.06875
125000 0.15625
130000 0.2625
135000 0.325
140000 0.40625
145000 0.43125
150000 0.43125
155000 0.41875
160000 0.44375
165000 0.45
170000 0.4375
175000 0.4625
180000 0.4625
185000 0.5125
190000 0.54375
195000 0.5625
200000 0.55625
205000 0.625
210000 0.61875
215000 0.63125
220000 0.70625
225000 0.725
230000 0.7
235000 0.71875
240000 0.7375
245000 0.7125
250000 0.76875
255000 0.80625
260000 0.8125
265000 0.8375
270000 0.8375
275000 0.825
280000 0.825
285000 0.81875
290000 0.76875
295000 0.76875
300000 0.775
305000 0.75625
310000 0.76875
315000 0.8
320000 0.76875
325000 0.7375
330000 0.7125
335000 0.69375
340000 0.6875
345000 0.73125
350000 0.71875
355000 0.74375
360000 0.76875
365000 0.7875
370000 0.7875
375000 0.8125
380000 0.7875
385000 0.78125
390000 0.76875
395000 0.775
400000 0.75
};
\addplot [thick, royalblue, dashed]
table {%
0 nan
5000 nan
10000 nan
15000 nan
20000 0
25000 0.01875
30000 0.0375
35000 0.04375
40000 0.04375
45000 0.04375
50000 0.04375
55000 0.0375
60000 0.06875
65000 0.0875
70000 0.13125
75000 0.1375
80000 0.18125
85000 0.175
90000 0.15625
95000 0.1125
100000 0.1375
105000 0.1625
110000 0.21875
115000 0.34375
120000 0.39375
125000 0.4
130000 0.36875
135000 0.35
140000 0.2625
145000 0.34375
150000 0.35625
155000 0.4
160000 0.36875
165000 0.36875
170000 0.2375
175000 0.2625
180000 0.20625
185000 0.21875
190000 0.25625
195000 0.35
200000 0.43125
205000 0.43125
210000 0.43125
215000 0.375
220000 0.425
225000 0.4125
230000 0.51875
235000 0.60625
240000 0.7125
245000 0.63125
250000 0.46875
255000 0.475
260000 0.3625
265000 0.35625
270000 0.3875
275000 0.4875
280000 0.45
285000 0.5375
290000 0.51875
295000 0.5875
300000 0.6375
305000 0.66875
310000 0.69375
315000 0.74375
320000 0.73125
325000 0.7125
330000 0.7125
335000 0.70625
340000 0.69375
345000 0.70625
350000 0.7125
355000 0.725
360000 0.75
365000 0.7625
370000 0.8
375000 0.83125
380000 0.83125
385000 0.81875
390000 0.83125
395000 0.64375
400000 0.50625
};
\end{axis}

\end{tikzpicture}